\documentclass[12pt]{article}
\usepackage{epsfig}
\newcommand{\mysection}{\setcounter{equation}{0}\section}

\def\beq{\begin{equation}}
\def\eeq{\end{equation}}
\def\beqa{\begin{eqnarray}}
\def\eeqa{\end{eqnarray}}

\newlength{\dinwidth} \newlength{\dinmargin}
\begin{document}
\begin {flushright}
Cavendish-HEP-03/22\\
LBNL-53701\\
\end {flushright} 
\vspace{3mm}
\begin{center}
{\Large \bf Top quark production at the Tevatron at NNLO\footnote{Presented
at the HEP2003 Europhysics Conference, Aachen, Germany, 17-23 July, 2003.}}
\end{center}
\vspace{2mm}
\begin{center}
{\large Nikolaos Kidonakis$^a$ and Ramona Vogt$^b$}\\
\vspace{2mm}
$^a${\it Cavendish Laboratory, University of Cambridge,\\
Madingley Road, Cambridge CB3 0HE, UK}\\
\vspace{2.5mm}
$^b${\it Nuclear Science Division,\\
    Lawrence Berkeley National Laboratory, Berkeley, CA 94720, USA \\
  and \\
  Physics Department,\\
    University of California at Davis, Davis, CA 95616, USA}
\end{center}

\begin{abstract}
We present results for top quark production at the Tevatron
including next-to-next-to-leading order (NNLO) soft-gluon 
corrections. We show the stability of the
cross section with respect to kinematics choice and
scale when the NNLO corrections are taken into account.
\end{abstract}

\thispagestyle{empty}  \newpage  \setcounter{page}{2}

\mysection{Introduction}
 
Improved theoretical calculations of the
top quark production cross sections and differential distributions
are becoming more important as the top is observed
at Tevatron Run II, where increasingly accurate
measurements of the top mass and cross section are expected.
The current state-of-the-art in the theory of top hadroproduction 
is calculations of next-to-next-to-leading order (NNLO) soft-gluon 
corrections to the double differential cross section \cite{KV,NKNNLO} 
from threshold resummation techniques \cite{KS,KOSr,LOS}. 
Near threshold there is limited phase space for the emission of real
gluons so that soft-gluon corrections dominate the cross section.

These soft corrections appear in the form of logarithmic ``plus''
distributions $[\ln^l(x_{\rm th})/x_{\rm th}]_+$, with $l \le 2n-1$ for 
the $n$-th order corrections in $\alpha_s$, where $x_{\rm th}$ is a kinematical
variable that measures distance from threshold.
Calculations for top quark production until recently had been done
through next-to-next-to-leading logarithmic (NNLL) accuracy, i.e.
including logarithms with $l=3,2,1$ at NNLO \cite{NKtop,KLMV}.
This NNLO-NNLL calculation  
greatly diminished the factorization and renormalization scale 
dependence of the cross section.
However, the dependence of the corrections on the choice of kinematics, 
single-particle-inclusive (1PI) or pair-invariant-mass (PIM), was substantial
even near threshold. 
Away from threshold, where hard real gluonic radiation becomes
non-negligible, the discrepancy between 1PI and PIM results is not surprising. 
However, near threshold the results should be the same if all the 
NNLO soft corrections are included.
Thus subleading (beyond NNLL) contributions can still have
an impact on the cross section.
 
We have recently calculated next-to-next-to-next-to-leading
logarithmic (NNNLL) terms  at NNLO in Ref. \cite{KV}, i.e. terms with $l=0$,
following the unified approach of Ref. \cite{NKNNLO}.
We find that the inclusion of NNNLL terms indeed eliminates 
the kinematics ambiguity near threshold.

\mysection{Analytical results}

We first discuss the analytical expressions for the corrections.
We begin with the next-to-leading order (NLO) corrections.  
In the ${\overline {\rm MS}}$ scheme, the NLO soft and virtual corrections 
for $q {\overline q} \rightarrow t {\overline t}$
in 1PI kinematics are
\beq
s^2\, \frac{d^2{\hat\sigma}^{(1)\; \rm 1PI}_{q {\overline q}}}{dt_1 \, du_1}
=F^{B\; \rm 1PI}_{q {\overline q}} 
\frac{\alpha_s(\mu_R^2)}{\pi} \left\{
c^{\rm 1PI}_{3 \;q {\overline q}} \left[\frac{\ln(s_4/m^2)}{s_4}\right]_+ 
+c^{\rm 1PI}_{2 \;q {\overline q}} \left[\frac{1}{s_4}\right]_+
+c^{\rm 1PI}_{1 \;q {\overline q}}  \, \delta(s_4) \right\}\, ,
\label{NLO1PI}
\eeq
where $F^{B\; \rm 1PI}_{q {\overline q}}$ is the Born term
and $\mu_R$ the renormalization scale. 
Also
$c^{\rm 1PI}_{3 \;q {\overline q}}=4C_F$ and
\beqa
c^{\rm 1PI}_{2 \;q {\overline q}}&=&C_A\left[-3\ln\left(\frac{u_1}{t_1}\right)
-\ln\left(\frac{m^2s}{t_1u_1}\right)+L'_{\beta}\right] 
\nonumber \\ && 
{}+2C_F\left[4\ln\left(\frac{u_1}{t_1}\right)
-\ln\left(\frac{t_1u_1}{m^4}\right)-L'_{\beta}-1
-\ln\left(\frac{\mu_F^2}{s}\right)\right]
\eeqa
where $\mu_F$ is the factorization scale,
$C_A=N_c=3$ is the number of colors, $C_F=(N_c^2-1)/(2N_c)$, and
$L'_{\beta}=[(1-2m^2/s)/\beta] \ln[(1-\beta)/(1+\beta)]$
with $\beta=\sqrt{1-4m^2/s}$.
For use below, we write 
$c^{\rm 1PI}_{2 \;q {\overline q}} \equiv T^{\rm 1PI}_{2 \;q {\overline q}}
-2C_F\ln(\mu_F^2/s)$. Finally,
$c^{\rm 1PI}_{1 \;q {\overline q}}=
\sigma^{(1)\, S+V\; \rm 1PI}_{q {\overline q} \, \delta}/
[(\alpha_s/\pi) F^{B\; \rm 1PI}_{q {\overline q}}]$
where $\sigma^{(1)\, S+V\; \rm 1PI}_{q {\overline q} \, \delta}$ 
denotes the $\delta(s_4)$ terms
in Eq. (4.7) of Ref. \cite{BNMSJ} with the definitions of $t_1$ and $u_1$ 
interchanged with respect to that reference.
We also define
$c^{\rm 1PI}_{1 \;q {\overline q}} \equiv T^{\rm 1PI}_{1 \;q {\overline q}}
+C_F [-3/2+\ln(t_1u_1/m^4)] \ln(\mu_F^2/s) 
+(\beta_0/2)\ln(\mu_R^2/s)$,
where $T^{\rm 1PI}_{1 \;q {\overline q}}$ has no scale dependence.

We further define the constants 
$\zeta_2=\pi^2/6$, and $\zeta_3=1.2020569\cdots$,
$\beta_0=(11C_A-2n_f)/3$,
and $K=C_A(67/18-\pi^2/6)-5n_f/9$ where $n_f$ is
the number of light quark flavors.

Following Ref. \cite{NKNNLO} we write the NNLO soft-plus-virtual 
corrections in 1PI kinematics as
\beqa
&& \hspace{-8mm}s^2\, \frac{d^2{\hat\sigma}^{(2)\; \rm 1PI}_{q {\overline q}}}
{dt_1 \, du_1}
=F^{B\; \rm 1PI}_{q {\overline q}} \frac{\alpha_s^2(\mu_R^2)}{\pi^2} 
\left\{\frac{1}{2} \left(c^{\rm 1PI}_{3 \;q {\overline q}}\right)^2 
\left[\frac{\ln^3(s_4/m^2)}{s_4}\right]_+
+\left[\frac{3}{2} c^{\rm 1PI}_{3 \;q {\overline q}} \, c^{\rm 1PI}_{2 \;q {\overline q}}
-\frac{\beta_0}{4} c_{3 \;q {\overline q}} ^{\rm 1PI} \right] 
\left[\frac{\ln^2(s_4/m^2)}{s_4}\right]_+
\right.
\nonumber \\ && \hspace{-2mm}
{}+\left[c^{\rm 1PI}_{3 \;q {\overline q}} \, c^{\rm 1PI}_{1 \;q {\overline q}}
+\left(c^{\rm 1PI}_{2 \;q {\overline q}}\right)^2
-\zeta_2 \, \left(c^{\rm 1PI}_{3 \;q {\overline q}}\right)^2
-\frac{\beta_0}{2} T^{\rm 1PI}_{2 \;q {\overline q}}
+\frac{\beta_0}{4} c_{3\;q {\overline q}}^{\rm 1PI} 
\ln\left(\frac{\mu_R^2}{s}\right) \right.
\nonumber \\ && \quad \left.
{}+2 C_F K+8 \frac{C_F}{C_A}\ln^2\left(\frac{u_1}{t_1}\right)\right] 
\left[\frac{\ln(s_4/m^2)}{s_4}\right]_+
\nonumber \\ && \hspace{-2mm}
{}+\left[c^{\rm 1PI}_{2 \;q {\overline q}} \, c^{\rm 1PI}_{1 \;q {\overline q}}
-\zeta_2 \, c^{\rm 1PI}_{2 \;q {\overline q}} \, c^{\rm 1PI}_{3 \;q {\overline q}}
+\zeta_3 \, \left(c^{\rm 1PI}_{3 \;q {\overline q}}\right)^2 
-\frac{\beta_0}{2} T_{1\;q {\overline q}}^{\rm 1PI} 
+\frac{\beta_0}{4} c_{2\;q {\overline q}}^{\rm 1PI} \ln\left(\frac{\mu_R^2}{s}\right)
+{\cal G}_{q{\overline q}}^{(2)}\right.
\nonumber \\ && \quad  \left.
{}+C_F\frac{\beta_0}{4} \ln^2\left(\frac{\mu_F^2}{s}\right)
-C_F K \ln\left(\frac{\mu_F^2}{s}\right)
-C_F K \ln\left(\frac{t_1 u_1}{m^4}\right)
+8 \frac{C_F}{C_A}\ln^2\left(\frac{u_1}{t_1}\right)
\ln\left(\frac{m^2}{s}\right) \right]
\left[\frac{1}{s_4}\right]_+
\nonumber \\ &&   \left.  
{}+R^{\rm 1PI}_{q {\overline q}} \delta(s_4)\right\} \, .
\label{NNLOqqMS}
\eeqa
Here the term
${\cal G}_{q \overline q}^{(2)}=C_F C_A 
(7 \zeta_3/2+22\zeta_2/3-299/27)+ n_f C_F(-4\zeta_2/3
+50/27)$
denotes a set of two-loop contributions universal for processes
with $q {\overline q}$ initial states \cite{NKNNLO}.  Process-dependent
two-loop corrections are not included in 
${\cal G}_{q \overline q}^{(2)}$ but their
contribution is expected to be negligible.
The virtual contribution $R^{\rm 1PI}_{q {\overline q}}$ is not fully known. 
However, we can determine certain terms in $R^{\rm 1PI}_{q {\overline q}}$
exactly.  These exact terms involve the factorization
and renormalization scales as well as the those terms ($\zeta$ terms)
that arise from the inversion from moment to momentum space \cite{KV}.

The analytical form of the NLO and NNLO corrections for the 
$q{\overline q}$ channel in PIM kinematics is similar,
see Ref. \cite{KV}. The expressions for the $gg$ channel
in both 1PI and PIM kinematics are more involved since the $gg$
color structure is more complex \cite{KV}.

\mysection{Numerical results}

\begin{figure}[htb] 
\setlength{\epsfxsize=1.0\textwidth}
\setlength{\epsfysize=0.5\textheight}
\centerline{\epsffile{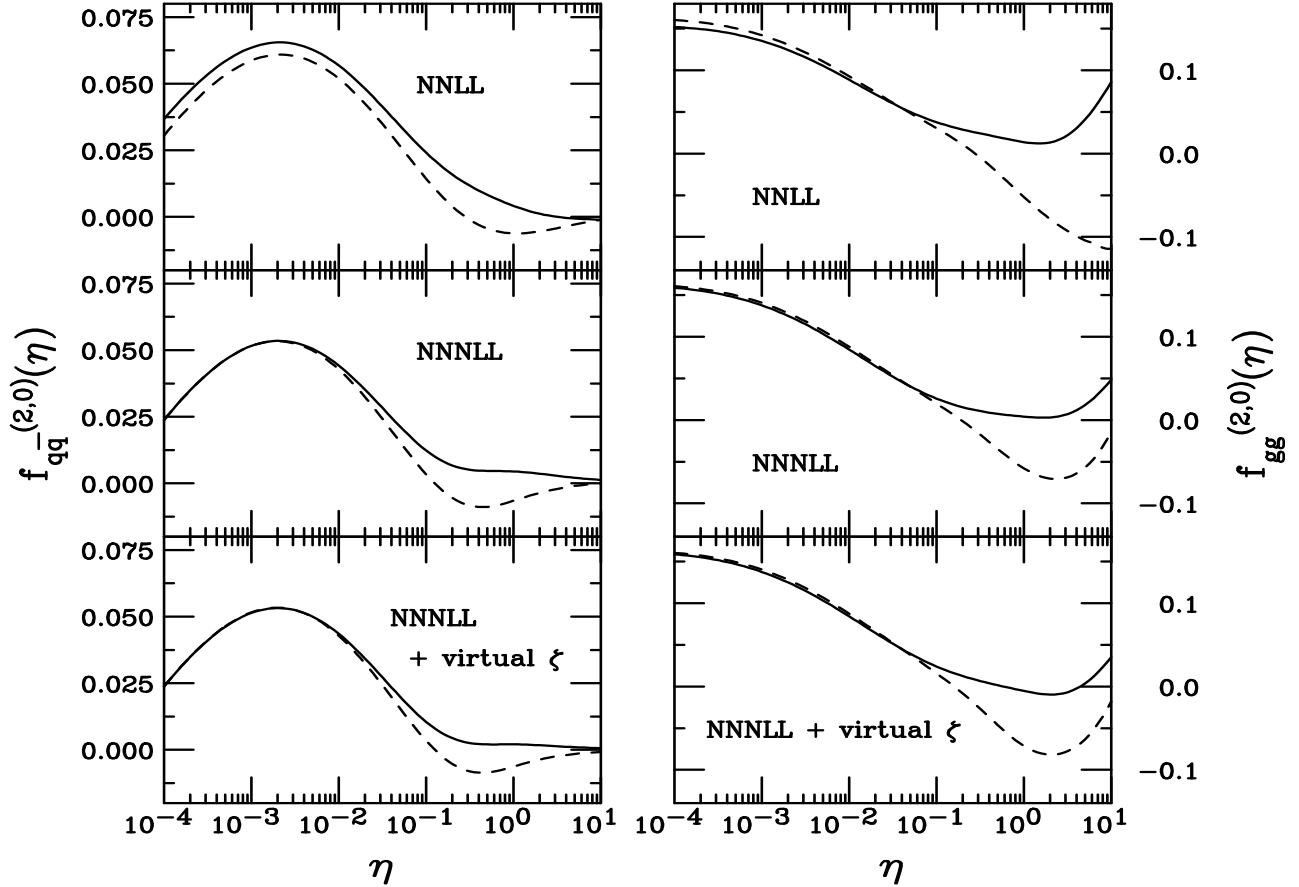}}
\caption[]{The $f_{ij}^{(2,0)}$ scaling functions in the
$\overline{\rm MS}$ scheme for the $q \overline q$ (left) 
and $gg$ (right) channels.  The top plots show the NNLL result, 
the center plots the NNNLL results and the bottom plots give the results 
including the virtual $\zeta$ terms.
The solid curves are for 1PI kinematics, the dashed for PIM kinematics.}
\label{fig:1} 
\end{figure}

We now study the numerical impact of the corrections.
The total partonic cross section may be expressed in terms of 
dimensionless scaling functions
$f^{(k,l)}_{ij}$ that depend only on $\eta \equiv s/4m^2 - 1$ \cite{KLMV},
\beq
\sigma_{ij}= \frac{\alpha^2_s(\mu)}{m^2}
\sum\limits_{k=0}^{\infty} \,\, \left( 4 \pi \alpha_s(\mu) \right)^k
\sum\limits_{l=0}^k \,\, f^{(k,l)}_{ij}(\eta) \,\,
\ln^l\left(\frac{\mu^2}{m^2}\right) \, .
\eeq
Here we have set $\mu \equiv \mu_F=\mu_R$.

In Fig.~\ref{fig:1} we plot the $f_{ij}^{(2,0)}$ scaling functions, 
the most important contributions at NNLO and independent of $\mu$.  
To demonstrate the effect of adding successive subleading contributions, 
we show the NNLL results in the upper plots, the scaling functions through 
NNNLL in the middle plots, 
and the results with the NNNLL and virtual $\zeta$ terms in the lower plots.  

We first discuss the results for the $q{\overline q}$ channel.
We note that to NNLL, the two kinematics choices give rather different results,
even at low $\eta$.  When the NNNLL terms are added the two results
coincide near threshold. Adding the
virtual $\zeta$ terms resulting from inversion improves the agreement 
between the 1PI and PIM kinematics further.  
This agreement also indicates that additional two-loop contributions 
not included in our expressions should be small.
A similar trend is seen for the $gg$ channel on the right-hand side of
Fig.~\ref{fig:1}.  The agreement between the NNLL 1PI and PIM scaling functions
at low $\eta$ is significantly better than in the $q \overline q$ channel.
Note however the significant divergence at large $\eta$.  
Again, inclusion of the subleading
contributions improves agreement over all $\eta$.  The improvement at 
larger $\eta$, $\eta > 0.1$ is notable.  
We remark that the effect of the virtual $\zeta$ terms is numerically 
small for both channels,  
in agreement with the arguments in Section IIIC of Ref.~\cite{NKtop}. 

\begin{figure}[htb] 
\setlength{\epsfxsize=1.0\textwidth}
\setlength{\epsfysize=0.3\textheight}
\centerline{\epsffile{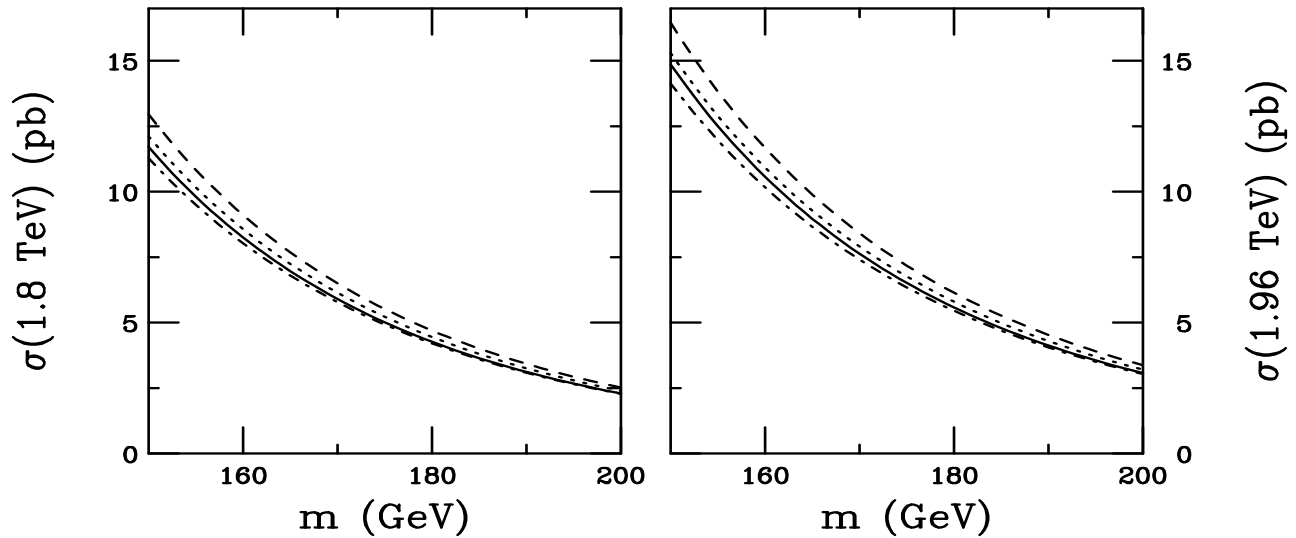}}
\caption[]{The $t \overline t$ total cross sections in $p \overline p$
collisions at $\sqrt{S} = 1.8$ TeV
(left-hand side) and 1.96 TeV (right-hand side) as functions of $m$ for
$\mu = m$.  The NLO (solid), and approximate NNLO 1PI (dashed), 
PIM (dot-dashed) and average (dotted) results are shown.
}
\label{fig:2} 
\end{figure}

We now turn to our calculations of the hadronic total cross sections.
We use the recent MRST2002 NNLO (approximate) parton
densities \cite{mrst2002} with an NNLO evaluation of $\alpha_s$.
Our calculations use the exact LO and NLO cross sections with the
soft NNNLL and virtual $\zeta$ corrections and the full soft-plus-virtual
scale-dependent terms at NNLO.  
In addition we multiply the NNLO scaling functions
by a damping factor,
$1/\sqrt{1+\eta}$, as in Ref.~\cite{KLMV}, to lessen the influence
of the large $\eta$ region where the threshold approximation does not
hold so well.

In Fig.~\ref{fig:2}, we present the NLO and approximate NNLO $t \overline t$
cross sections at $\sqrt{S} = 1.8$ TeV (left-hand side) and 1.96 TeV
(right-hand side) as functions of top quark mass for $\mu = m$.   
We also show the average of the two kinematics results.
Both the 1PI and PIM cross sections are reduced
due to the subleading terms.  The average of the two
kinematics results is just above the NLO cross sections for both energies.

Our results using the CTEQ6M NLO parton densities \cite{CTEQ6}
are similar. At $\sqrt{S} = 1.8$ TeV, averaging over the NNLO 1PI and PIM 
results with the two sets of parton distributions at $\mu=m=175$ GeV, 
our best estimate for the cross section is
$5.24 \pm 0.31$ pb where the quoted uncertainty is due to the kinematics
choice. At $\sqrt{S} = 1.96$ TeV our corresponding best estimate is
$6.77 \pm 0.42$ pb. We note that differences with our previous
estimates in \cite{NKtop,KLMV} have as much to do with the use of
the new parton densities 
as with the inclusion of the new subleading terms.
We also note that the scale dependence of the NNLO cross section is negligible
\cite{KV}.

\mysection*{Acknowledgements}

The research of N.K. has been supported by a Marie Curie Fellowship of 
the European Community programme ``Improving Human Research Potential'' 
under contract number HPMF-CT-2001-01221.
The research of R.V. is supported in part by the 
Division of Nuclear Physics of the Office of High Energy and Nuclear Physics
of the U.S. Department of Energy under Contract No. DE-AC-03-76SF00098.

\end{document}